\documentclass{ifacconf}

\usepackage{graphicx}      
\usepackage{natbib}        

\usepackage{amsmath}
\usepackage{amssymb,graphicx}
\usepackage{epstopdf}
\usepackage{mathtools}
\usepackage{algorithm}
\usepackage[noend]{algpseudocode}
\usepackage{xcolor}
\usepackage{subcaption}

\newcommand{\red}[1]{{\color{red}#1\color{black}} } 

\newtheorem{example}{Example}

\makeatletter
\def\BState{\State\hskip-\ALG@thistlm}
\makeatother
\algnewcommand\algorithmicinput{\textbf{Input:}}
\algnewcommand\INPUT{\item[\algorithmicinput]}
\algnewcommand\algorithmicoutput{\textbf{Output:}}
\algnewcommand\OUTPUT{\item[\algorithmicoutput]}

\newcommand{\DTDS}{dt-DS }

\newcommand{\R}{\mathbb{R}}
\newcommand{\N}{\mathbb{N}}

\newcommand{\im}{\operatorname{im}}
\newcommand{\ind}{\operatorname{ind}}

\newcommand{\Sys}{\Sigma} 
	\newcommand{\T}{{\mathbb{T}}} 
	\newcommand{\W}{{\mathbb{W}}} 
	\newcommand{\Beh}{{\mathfrak{B}}} 
	
	\newcommand{\Biso}{\Beh_{\mbox{\small i/s/o}}}
	\newcommand{\Bio}{\Beh_{\mbox{\small i/o}}}
\newcommand{\SysX}{\Sigma_{\mathcal{C}}} 

\newcommand{\M}{\mathbf{M}}
\newcommand{\X}{\mathbb{X}}
\newcommand{\Y}{\mathbb{Y}}
\newcommand{\A}{\mathbb{U}}
\newcommand{\B}{\mathbb{S}}

\newcommand{\re}{{\mathcal{R}}}
\newcommand{\I}{{\mathcal{I}}}
\newcommand{\F}{{\mathcal{F}}}
\newcommand{\C}{{\mathfrak{C}}}

\newcommand{\Sa}{{\mathcal{S}_a}}

\newcommand{\DV}{\Sys_{\mbox{\scriptsize{DV}}}}
\newcommand{\DVa}{\Sys_{\mbox{\scriptsize{DV}}_a}}

\newcommand{\redx}[1]{}
\usepackage{enumerate}
\usepackage{sidecap}

\begin{document}

\begin{frontmatter}

\title{Control refinement for discrete-time descriptor systems: a behavioural approach via simulation relations} 


\author[TUE]{F. Chen},  %
\author[TUE]{S. Haesaert}, %
\author[OX]{A. Abate}, and  %
\author[TUE]{S. Weiland} 

 \address[TUE]{Department of Electrical Engineering\\ Eindhoven University of Technology, 
   Eindhoven, The Netherlands} 
\address[OX]{Department of Computer Science\\ University of Oxford, Oxford, United Kingdom }

\begin{abstract}                
The analysis 
 of industrial processes, modelled as descriptor systems, is often computationally hard due to the presence of both algebraic couplings and difference equations of high order. In this paper, we introduce a control refinement notion 
 for these descriptor systems 
 that enables analysis and control design over related reduced-order systems. Utilising the behavioural framework, we extend upon 
 the standard hierarchical control refinement for ordinary systems  
  and allow for algebraic couplings inherent to descriptor systems. 
\end{abstract}

\end{frontmatter}
\section{Introduction}%
%
\vspace{-1mm}
Complex industrial processes generally 
contain algebraic couplings in addition to differential (or difference) equations of high order.
These systems, referred to as descriptor systems \citep{kunkel2006differential,dai1989singular}, are commonly used in the modelling of mechanical systems. 
The presence of algebraic equations, or couplings, together with large state dimensions renders numerical simulation and controller design  challenging. 
%
Instead model reduction methods 
\citep{antoulas2005approximation} can be applied to   replace the systems with reduced order ones.  Even though most methods have been developed for systems with only ordinary difference equations, recent research also targets descriptor systems  \citep{cao2015hankel}. \\*
In this paper, we newly target the use of descriptor systems of reduced order for the verifiable design of controllers. 
A rich body of literature on verification and formal controller synthesis exists for systems solely composed of difference equations.  
This includes the algorithmic design of certifiable (hybrid) controllers and the verification of pre-specified requirements~\citep{tabuada,kloetzer2008fully}. 
Usually, these methods first reduce the original, concrete systems to abstract systems with finite or smaller dimensional state spaces 
over which the verification or controller synthesis can be run. 
A such controller obtained for the abstract system can be refined over the concrete system leveraging the existence of a similarity relation, e.g.,  an (approximate) simulation relation, between the two systems \citep{tabuada, bridge}.  
%
For the application of these relations in control problems,  
a hierarchical control framework 
is presented by~\citep{main}. 
Currently, the control synthesis over descriptor systems cannot be dealt with in this fashion due to the presence of algebraic equations. 


The presence of similarity relations between descriptor systems has also been a topic under investigation in~\citep{megawati2015bisimulation}.
This work on similarity relations deals  with continuous-time descriptor systems that are unconstrained and non-deterministic, 
and focuses on the conditions for bisimilarity and on the construction of similarity relations. 
Instead in this work, we specifically consider the control refinement problem for discrete-time descriptor systems via simulation relations within a behavioural framework, 
such that properties verified over the future behaviour of the abstract system are also verified over the concrete controlled system. Within the behavioural theory \citep{willems2013introduction}, 
 a formal distinction is made between a system (its behaviour) and its representations, enabling  us to  
 investigate descriptor systems and refinement control problems without having to directly deal with their inherent anti-causality.\\*
 In the next section, we define the notion of dynamical systems and control within a behavioural framework and use it to formalise the control refinement  problem. Subsequently, Section \ref{sec:Main} is dedicated to the exact control refinement for descriptor systems and contains the main results of the paper. The last section closes with the conclusions.  

\section{The behavioural framework}
\subsection{Discrete-time descriptor systems}
As introduced by \citep{willems2013introduction},  we define dynamical systems as follows.
\begin{defn}\label{def:DynSys}
A dynamical system $\Sys$ is defined as a triple
\begin{equation*}
\Sys=(\T,\W,\Beh)
\end{equation*} with
  the \emph{time axis}  $\T$, the \emph{signal space} $\W$, and the \emph{behaviour} $\Beh\subset\W^\T$. \qed  
\end{defn}
In this definition, $\W^\T$ denotes the collection of all time-dependent functions 
$w:\T\rightarrow \W$. 
The set of trajectories or time-dependent functions given by $\Beh$ represents the trajectories that are compatible with the system. This set is referred to as the behaviour of the system \citep{willems2013introduction}. Generally, the representation of the behaviour of a dynamical system by equations, 
such as a set of ordinary
differential equations, state space equations and transfer functions, is non-unique. 
Hence we distinguish a dynamical system (its behaviour) from the mathematical equations used to represent its governing laws. 

We consider dynamical systems evolving over discrete-time ($\T:=\N=\{0,1,2,\ldots\}$) that can be represented by a combination of linear difference and algebraic equations. 
%
%
The dynamics of such a linear discrete-time descriptor system (DS) are defined by the tuple $(E,A,B,C)$ as
\begin{equation}\label{eq:DTDS}
\begin{aligned}
Ex(t+1)&=Ax(t)+Bu(t),\\
y(t)&=Cx(t),\hspace{.5cm} 
\end{aligned} 
\end{equation}
with the state  $x(t) \in\X = \R^n$, the input $u(t)\in\A =  \R^p$, and the output $y(t)\in\Y = \R^k$ and $t\in \N$. Further, $E,A\in \R^{n\times n},B\in \R^{n\times p}$ and $C\in \R^{k\times n}$ are constant matrices and we presume that rank$(B)=p$ and rank$(C)=k$. 

We say that a trajectory $w=(u,x,y)$, with $w:\N\rightarrow (\A\times\X\times\Y)$, satisfies  \eqref{eq:DTDS}  if for all $t\in \N$  the equations in \eqref{eq:DTDS} evaluated at  $u(t), x(t),x(t+1),y(t)$ hold.
 Then the collection of all trajectories $w$ defines the \emph{full} behaviour,  or equivalently the \emph{input-state-output} behaviour as \begin{equation}\label{eq:Beh_iso}
 \Biso :=\{(u,x,y)\in (\A\times \X\times \Y)^\N\mid (\ref{eq:DTDS}) \mbox{ is satisfied}\}.\end{equation}
The variable $x$ is considered as a latent variable, therefore the \emph{manifest}, or equivalently the \emph{input-output} behaviour associated with \eqref{eq:DTDS} is defined by
\begin{equation*}
\begin{aligned}
\Bio \!\! :=\{(u,y){\mid} \exists &x\in \X^\N  \,\,\mbox{s.t.}\,\,(u,x,y)\in \Beh_{\mbox{i/s/o}}\}.
\end{aligned}
\end{equation*}
If $E$ is non-singular, we refer to the corresponding dynamical system as  a \emph{non-singular DS}. In that case, we can transform \eqref{eq:DTDS} into  standard state space equations, as   
\begin{equation} \label{eq:stdtDS}
\begin{aligned}
x(t+1)&=\tilde Ax(t)+\tilde Bu(t), \\
y(t)&=Cx(t),\end{aligned}
\end{equation}
with $\tilde A=E^{-1}A , \tilde B=E^{-1}B$. Further $\Biso$ as in \eqref{eq:Beh_iso} is  $$\{(u,x,y)\in (\A\times \X\times \Y)^\N\mid (u,x,y) \mbox{ s.t. }\eqref{eq:stdtDS}\mbox{ holds}\}.$$ 
Similarly, if $E$ is non-singular, $\Bio$ can be defined by \eqref{eq:stdtDS}.

The tuple with dynamics \eqref{eq:DTDS} defines a dynamical system $\Sys$ evolving over the combined signal space $\W= \A\times\X\times\Y$ with behaviour $\Beh:=\Biso$ given in \eqref{eq:Beh_iso}. Similarly, for $\W$ restricted to input-output space, the tuple $(\N, \A\times\Y, \Bio)$ defines the manifest or induced dynamical system.  
 
We are specifically interested in the behaviour  initialised at $t=0$ with a given set of initial states $ \X_0\subset \X$. 
For this,  we say that a 
trajectory $w:\N\rightarrow (\A\times\X\times\Y)$ is initialised with $\X_0$ if \eqref{eq:DTDS} holds and $x(0)=x_0\in\X_0$. Such a trajectory, initialised with $x_0\in \X_0$, is also called the continuation of $x_0$.
We refer to the collection of initialised trajectories related to $\X_0$ as the initialised behaviour $\Biso^{init}$.
This allows us to formalise our definition of the descriptor system evolving over $\N$.
\begin{defn}[Discrete-time descriptor systems (DS)]\mbox{ }
A (discrete-time) descriptor system is defined as a dynamical system $\Sys$ initialised with $\X_0$, whose behaviour can be represented  by the combination of algebraic equations and difference equations given in \eqref{eq:DTDS}, that  is
\begin{align}\label{eq:Sysdyn}\Sys:=\left(\T, \W,\Beh \right)=(\N,\A\times \X\times \Y,\Biso^{init})\end{align}
with
\begin{itemize}
\item the time axis $\T:=\N=\{0,1,2,\ldots\}$,  
\item the full signal space $\W := \A\times \X\times \Y, $ and
\item the \emph{initialised} behaviour\footnote{In the sequel the indexes $init$ and $i/s/o$ will be dropped.}
\begin{align*}\Biso^{init}=\{w\in\W^{\N}|w=(u,x,y) \mbox{ s.t. } \eqref{eq:DTDS}\hspace{2cm} \\ \hfill \mbox{ and s.t. } x(0)=x_0\in\X_0 \}.\end{align*}
\end{itemize}
\end{defn}
\subsection{Control of descriptor systems}
\begin{figure}[h]
\begin{subfigure}[b]{.5\columnwidth}\centering
\includegraphics[scale=.48]
{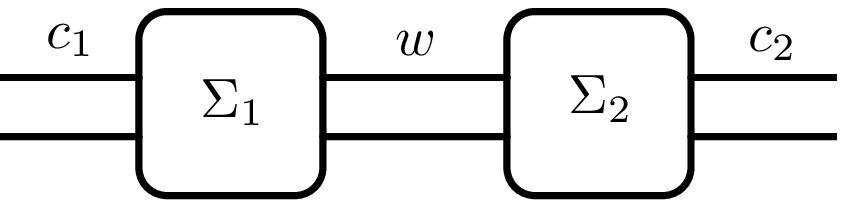}\vskip.1cm
\caption{The interconnected system $\Sys$ obtained via the shared variables $w$ in $\W$ between dynamical systems $\Sys_1$ and $\Sys_2$ with  signal spaces $\mathbb{C}_1\times\W$ and $\mathbb{C}_2\times \W$.}\label{fig:partialstructure}
\end{subfigure}\hfill
\begin{subfigure}[b]{.48\columnwidth}\centering
\includegraphics[scale=.5]
{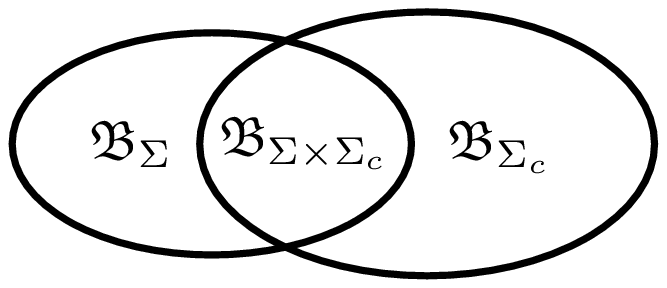}
\caption{The controlled behaviour $\Beh_{\Sys\times \Sys_c}=\Beh_{\Sys}\cap\Beh_{\Sys_c}$ is given as  the intersection of the behaviours of the dynamical system $\Sys$ and its controller $\Sys_c$. }\label{fig:cb}
\end{subfigure}
\caption{The left figure (a) portrays the general  interconnection of two dynamical systems. In figure (b), the more specific case of behavioural intersection for a system  and its controller is depicted. }
\end{figure}

Controller synthesis amounts to synthesising a system $\Sys_c$, called a controller, which, after interconnection with $\Sys$, restricts the behaviour $\Beh$ of $\Sys$ to desirable (or controlled) trajectories.  
Thus, in the behavioural framework,  control is defined through \emph{interconnections} (or via variable sharing as specified next), 
rather than based on  the causal transmission of  signals or information, as in classical system theory. Let $\Sys_1=(\T,\mathbb{C}_1\times \W,\Beh_1)$ and $\Sys_2=(\T,\mathbb{C}_2\times \W,\Beh_2)$ be two dynamical systems. Then, as depicted in  Fig. \ref{fig:partialstructure} and defined in \citep{willems2013introduction}, the \emph{interconnection} of $\Sys_1$ and $\Sys_2$ over $\W$, denoted by $\Sys=\Sys_1\times_w \Sys_2$ with the shared variable $w\in \W$, yields the dynamical system $\Sys=(\T,\mathbb{C}_1\times \mathbb{C}_2\times \W,\Beh)$ with $\Beh=\{(c_1,c_2,w):\T\rightarrow \mathbb{C}_1\times \mathbb{C}_2\times \W\mid (c_1,w)\in \Beh_1,(c_2,w)\in \Beh_2\}$.

Observe that $w\in \W^\T$ contains the signals shared by both $\Sys_1$ and $\Sys_2$, while $c_1\in \mathbb{C}_1^\T$ only belongs to $\Sys_1$ and $c_2\in \mathbb{C}_2^\T$ only belongs to $\Sys_2$. So, in the interconnected system, the shared variable $w$ satisfies the laws of both $\Beh_1$ and $\Beh_2$. Note that it is always possible to trivially extend the signal spaces of $\Sys_1$ and $\Sys_2$ (and the associated behaviour)  such that a full interconnection structure is obtained, that is, such that both $\mathbb{C}_1$ and $\mathbb{C}_2$ are empty and the behaviour of the interconnected system is $\Beh=\Beh_1\cap \Beh_2$. Hence, a full interconnection of $\Sys=(\T,\W,\Beh_\Sys)$ and $\Sys_c=(\T,\W,\Beh_{\Sys_c})$ is simply $\Sys\times_w \Sys_c=(\T,\W,\Beh_{\Sys}\cap \Beh_{\Sys_c})$, with the intersection of the behaviours, denoted by $\Beh_{\Sys\times \Sys_c}$, as portrayed in  Fig. \ref{fig:cb}. That is, interconnection and intersection are equivalent in full interconnections.


Further, we  define a well-posed controller $\Sys_c$ for $\Sys$ as follows.  

\begin{defn}\label{well-posed}
Consider a dynamical system $\Sys=(\T,\W,\Beh)$, with initialised behaviour as defined in \eqref{eq:Sysdyn}.
%
 We say that a system $\Sys_c=(\T,\W,\Beh_c)$ is a \emph{well-posed} controller for $\Sys$ if the following conditions are satisfied:
\begin{enumerate}
\item $\Beh_{\Sys\times \Sys_c}:=\Beh_\Sys\cap \Beh_{\Sys_c}\neq \{\emptyset\};$
\item For every initial state $x_0\in \X_0$, there exists a unique continuation in $\Beh_{\Sys\times \Sys_c}$.
\end{enumerate}
Denote with $\C(\Sys)$ the collection of all well-posed controllers for $\Sys$.
\end{defn}


We want a controller that  accepts any initial state of the system. This is formalised in the second condition by requiring that for any initial state of $\Sys$, there exists a unique continuation in $\Beh_{\Sys\times\Sys_c}$.
We elucidate the properties of a well-posed linear controller as follows.
\begin{example}\label{linearwellposed}
\it
For a system $\Sys$ as in \eqref{eq:DTDS}, consider a controller $\Sys_{c}$, which is a DS, and has dynamics given as
\begin{equation}\label{DAEcontroller}
E_cx(t+1)=A_cx(t)+B_cu(t),
\end{equation}
with $E_c,A_c\in \R^{n_c\times n}$ and $B_c\in \R^{n_c\times p}$. Suppose that the controller shares the variables $u$ and $x$ with the system $\Sys$. That is, $w=(u,x)$. 
The interconnected system $\Sys \times_w \Sys_{c}$ yields the state evolutions of the combined system as
\begin{equation}\label{eq:DAEcontrollerinter}
\begin{bmatrix}E\\E_c\end{bmatrix}x(t+1)=\begin{bmatrix}A\\A_c\end{bmatrix}x(t)+\begin{bmatrix}B\\B_c\end{bmatrix}u(t),
\end{equation}
and can be rewritten to 
\begin{equation}\label{DAErewrite11}
\begin{bmatrix}E&-B\\E_c&-B_c\end{bmatrix}\begin{bmatrix}x(t+1)\\u(t)\end{bmatrix}=\begin{bmatrix}A\\A_c\end{bmatrix}x(t).
\end{equation}

If for any $x(t)\in \X$, there exists a pair $(x(t+1),u(t))$ such that (\ref{DAErewrite11}) holds, then this implies that for any initial state $x_0\in\X_0$ of $\Sys$ there exists a continuation in the controlled behaviour. In addition, if the pair $(x(t+1),u(t))$ is unique for any $x(t)\in\X$, 
then this continuation is unique and we say that $\Sys_{c}\in \C(\Sys)$. This existence and uniqueness of the pairs $(x(t+1),u(t))$ depends on the solutions of the matrix equality (\ref{DAErewrite11}).  We use the classical results on the solutions of matrix equalities (cf.  \citep{abadir2005matrix}) to  conclude that the first well-posedness condition is satisfied if and only if 
\begin{equation}\label{lemma9}
\operatorname{rank}\begin{pmatrix}
\begin{bmatrix}
E&B\\E_c&B_c
\end{bmatrix}
\end{pmatrix}
=\operatorname{rank}\begin{pmatrix}
\begin{bmatrix}
E&B&A\\E_c&B_c&A_c
\end{bmatrix}
\end{pmatrix}. 
\end{equation}
If in addition,  
\begin{equation}\label{lemma99} \operatorname{rank}\begin{pmatrix}
\begin{bmatrix}
E&B&A\\E_c&B_c&A_c
\end{bmatrix}
\end{pmatrix}= n+p, 
\end{equation}  then the second condition is also satisfied and $\Sys_{c}\in \C(\Sys)$.
\end{example}

 Of interest is the design of well-posed controllers subject to specifications over the future output behaviour of the  controlled system.  
We thus consider specifications defined over the output space. 
In order to analyse the output behaviour, we introduce a projection map. 
For $\Beh\subset (\W_1\times\W_2)^\T$ we denote with $\Pi_{\W_2}$ a projection given as  
\[ \Pi_{\W_2}\left(\Beh\right):=\{w_2\in\W_2^\T|\exists w_1\in\W_1^\T \mbox{ s.t. } (w_1,w_2)\in\Beh\}. \]
 
We focus here on finding a controller $\Sys_c$ for a given dynamical
system $\Sys$ such that the output behaviour $\Pi_\Y(\Beh_{\Sys\times \Sys_c})$ of the interconnected system satisfies some specifications.
 
\subsection{Exact control refinement \& problem statement}
Let us refer to the original DS that represents the real physical system as the \emph{concrete} DS. It is for this system that we would like to develop a well-posed controller.  
 Recall that the DS is a dynamical system $\Sys$ with dynamics $(E,A,B,C)$ as in (\ref{eq:DTDS}) and initialised with $\X_0$. 
A well-posed  controller  for $\Sys$ is referred to $\Sys_c\in\C(\Sys)$. The controlled concrete system is the interconnected system $\Sys\times_w \Sys_c$ with the shared variables $w=(u,x)$. 


Now, we consider a  simpler DS $\Sys_a$, related to the concrete DS $\Sys$, with dynamics given as $(E_a, A_a, B_a, C_a)$ and initialised with $\X_{a0}$. We assume that the synthesis of a well-posed controller $\Sys_{c_a}$ for  $\Sys_a$ is substantially easier than  for $\Sys$. We refer to this simpler system $\Sys_a$ as the \emph{abstract} DS, and we note that its signals take values $u_a(t),x_a(t),y_a(t)$ with   $x_a(t)\in \X_a = \R^m, u_a(t)\in \A_a = \R^q, y_a(t)\in \Y_a =\Y= \R^k$ and $t\in \N$. With respect to the concrete system, the abstract DS is generally a reduced-order system. The controlled abstract system $\Sys_a\times_{w_a} \Sys_{c_a}$ is the interconnected system with the shared variables $w_a=(u_a,x_a)$.
 
If we assume that we can compute a well-posed controller for the abstract system, then the control synthesis problem reduces to a control refinement problem. 
\begin{defn}[Exact control refinement]\label{def:exact_refine}
Let $\Sys_a$ and $\Sys$ be the abstract and concrete DS, respectively. We say that controller $\Sys_{c}\in\C(\Sys)$ refines the controller $\Sys_{c_a}\in \C(\Sys_a)$ if  $\Pi_\Y(\Beh_{\Sys\times \Sys_c})\subseteq \Pi_\Y(\Beh_{\Sys_a\times \Sys_{c_a}})$.
\end{defn}\vspace{-2mm}
Then we formalise the exact control refinement problem.\vspace{-2mm}
\subsubsection{Problem 1.}
Let $\Sys_a$ and $\Sys$ be the abstract and concrete DS, respectively. For any $\Sys_{c_a}\in \C(\Sys_a)$, refine $\Sys_{c_a}$ to $\Sys_c$, s.t. $\Sys_c\in \C(\Sys)$ and $\Pi_\Y(\Beh_{\Sys\times \Sys_c})\subseteq \Pi_\Y(\Beh_{\Sys_a\times \Sys_{c_a}})$.
\smallskip \vspace{-2mm}

 In the next section, we will show that the existence of a solution to this problem hinges on certain conditions involving similarity relations between the concrete and abstract DS.
For this, we will first introduce simulation relations to formally characterise this similarity.

\section{Exact control refinement}\label{sec:Main}


\subsection{Similarity relations between DS}

We give the notion of simulation relation as defined in \citep{tabuada} for transition systems and applied to pairs of DS $\Sys_1$ and $\Sys_2$ that share the same output space $\Y_1=\Y_2=\Y$.

\smallskip

\begin{defn} \label{Def:simrelation}
Let $\Sys_1$ and $\Sys_2$ be two DS with respective  dynamics $(E_1,A_1,B_1,C_1)$ and $(E_2,A_2,B_2,C_2)$ over state spaces $\X_1$ and $\X_2$. A relation $\re \subseteq \X_1\times \X_2$ is called a \emph{simulation relation from $\Sys_1$ to $\Sys_2$}, if $\forall (x_1,x_2)\in \re$,  
\begin{enumerate}
\item for all $(u_1,x_1^+)\in \A_1\times \X_1$ subject to \[E_1x_1^+=A_1x_1+B_1u_1\]   there exists  $(u_2,x_2^+)\in \A_2\times \X_2$ subject to \[E_2x_2^+=A_2x_2+B_2u_2\] such that $(x_1^+,x_2^+)\in \re$, and \\
\item we have $C_1 x_1=C_2 x_2$.
\end{enumerate}
We say that $\Sys_1$ is simulated by $\Sys_2$, denoted by $\Sys_1\preceq \Sys_2$, if there exists a simulation relation $\re$ from $\Sys_1$ to $\Sys_2$ and if in addition $\forall x_{10}\in \X_{10}, \exists x_{20}\in \X_{20}$ such that $(x_{10},x_{20})\in \re$.
\end{defn}
We call $\re\subseteq \X_1\times \X_2$ a \emph{bisimulation relation} between $\Sys_1$ and $\Sys_2$,
 if $\re$ is a simulation relation from $\Sys_1$ to $\Sys_2$ and its inverse $\re^{-1}\subseteq \X_2\times \X_1$ is a simulation relation from $\Sys_2$ to $\Sys_1$.  
 We say that $\Sys_1$ and $\Sys_2$ are bisimilar, denoted by $\Sys_1\cong \Sys_2$, if $\Sys_1\preceq\Sys_2$ w.r.t. $\re$ and $\Sys_2\preceq\Sys_1$ w.r.t. $\re^{-1}$. 
 
Simulation relations as defined above are transitive. Let $\re_{12}$ and $\re_{23}$ be simulation relations respectively, from $\Sys_1$ to $\Sys_2$ and from $\Sys_2$ to $\Sys_3$. Then a simulation relation from $\Sys_1$ to $\Sys_3$ is given as a composition of $\re_{12}$ and $\re_{23}$, namely 
\[\re_{12}\circ\re_{23}\!=\!\{(x_1,x_3)\mid \exists x_2\!:\! (x_1,x_2)\in \re_{12}\wedge(x_2,x_3)\in \re_{23}\}.\]
We also have that $\Sys_1\preceq\Sys_2$ and $\Sys_2\preceq\Sys_3$ implies $\Sys_1\preceq \Sys_3$ and, in addition, $\Sys_1\cong\Sys_2$ and $\Sys_2\cong\Sys_3$ implies $\Sys_1\cong \Sys_3$.

Simulation relations have also implications on the properties of the output behaviours of the two systems. More precisely, if a system is simulated by another system then this implies output behaviour inclusion. This follows from Proposition 4.9 in \citep{tabuada} and is formalised next.
\begin{prop}\label{Def:behimplication}
Let $\Sys_1$ and $\Sys_2$ be two DS with simulation relations as defined in Definition \ref{Def:simrelation}. Then,
\begin{equation}\nonumber
\begin{aligned}
&\Sys_1\preceq \Sys_2 \Longrightarrow \Pi_{\Y}(\Beh_{\Sys_1})\subseteq \Pi_{\Y}(\Beh_{\Sys_2}),\\
&\Sys_1\cong \Sys_2 \Longrightarrow \Pi_{\Y}(\Beh_{\Sys_1})=\Pi_{\Y}(\Beh_{\Sys_2}).
\end{aligned}
\end{equation}
\end{prop}\vspace{-2mm}
 \color{black}
Simulation relations can also be used for the controller design for deterministic systems such as nonsingular DS \citep{tabuada,fainekos2007hierarchical,main}.
This will be used in the next subsection, where we consider the exact control refinement for non-singular DS. After that, we introduce a transformation of a singular DS to an auxiliary nonsingular DS representation, referred to as a driving variable (DV) system. The exact control refinement problem is then solved based on the introduced notions.

\subsection{Control refinement for non-singular DS}
 Let us consider the simple case where the concrete and abstract systems of interest are given with non-singular dynamics. 
For these systems, the existence of a simulation relation also implies the existence of an interface function \citep{main}, which is formulated as follows.  
\begin{defn}\label{def:interface}
(Interface). Let $\Sys_1$ and $\Sys_2$ be two non-singular DS defined over the same output space $\Y$ with a simulation relation $\re$  from $\Sys_1$ to $\Sys_2$. A mapping $\F:\A_1\times \X_1\times \X_2\mapsto \A_2$ is an \emph{interface related to $\re$}, if  $\forall (x_1,x_2)\in\re$ and for all $u_1\in\A_1$, $u_2:=\F(u_1,x_1,x_2)\in \A_2$ is such that $(x_1^+,x_2^+)\in \re$ with 
 \[x_1^+=A_1x_1+B_1u_1 \mbox{ and }x_2^+=A_2x_2+B_2u_2.\]
%
\end{defn}\vspace{-2mm}
It follows from Definition \ref{Def:simrelation} that there exists at least one interface related to $\mathcal{R}$ if two deterministic, or non-singular systems are in a simulation relation. As such we can solve the exact refinement problem  as follows. 
%

\begin{thm}\label{standarddaeexact}
Let  $\Sys_1$ and $\Sys_2$ be two non-singular DS defined over the same output space $\Y$  with dynamics $(I,A_1,B_1,C_1)$ and $(I,A_2,B_2,C_2)$, which are initialised with $\X_{10}$ and $\X_{20}$, respectively. If there exists a relation $\re\subseteq \X_1\times\X_2$ such that \begin{enumerate}
\item $\re$ is a simulation relation from $\Sys_1$ to $\Sys_2$, and 
\item $\forall x_{20}\in \X_{20}, \exists x_{10}\in \X_{10}$ s.t. $(x_{10},x_{20})\in \re$, 
\end{enumerate}
then for any  controller $\Sys_{c_1} \in \C(\Sys_1)$, there exists a controller $\Sys_{c_2} \in \C(\Sys_{2})$ that is an exact control refinement for $\Sys_{c_1}$ and thus achieves with  
\[\Pi_\Y(\Beh_{\Sys_{2}\times \Sys_{c_2}})\subseteq \Pi_\Y(\Beh_{\Sys_1\times \Sys_{c_1}}).\]
 \end{thm}

\begin{pf}
Since $\re$ is a simulation relation  from $\Sys_1$ to $\Sys_2$, there exists an interface function $\F:\A_1\times\X_1\times\X_2\rightarrow \A_2$ as given in Definition \ref{def:interface}, cf \citep{tabuada,main}.
Additionally, due to (2) there exists a map, $\F_{0}:\X_{20}\rightarrow \X_{10}$ such that for all $x_{20}\in \X_{20}$ it holds that $(\F_{0}(x_{20}), x_{20})\in \re$.\\
Next, we construct the controller $\Sys_{c_2}$ that achieves exact control refinement for $\Sys_{c_1}$ as \[\Sys_{c_2}:=(\Sys_{1}\times_{w_1} \Sys_{c_1})\times_{w_1} \Sys_{\F},\]
where $w_1=(u_1,x_1)$ and where $\Sys_\F:=(\N,\W,\Beh_\F)$ is a dynamical system taking values in the combined signal space with 
\begin{align*}\Beh_\F:=\{(x_1,u_1,x_2,u_2)\in \W{\mid} x_{10}=\F_0(x_{20}) \mbox{ and }\hspace{1cm} \\  \hspace{2cm} u_2=\F(x_1,u_1,x_2)\}.\end{align*}
The dynamical system 
$\Sys_{c_2}$ is a well-posed controller for $\Sys_2$ with $\Sys_2\times_{w_2} \Sys_{c_2}$ sharing $w_2=(u_2,x_2)$. Denote with $\Beh_{\Sys_{2}\times\Sys_{c_2}}$ the behaviour of the controlled system, then due to the construction of $ \Sys_{\F}$ it follows that $\Beh_{\Sys_{2}\times\Sys_{c_2}}$ is non-empty and $\forall x_{20}\in \X_{20},\,\exists x_{10}\in \X_{10}$ such that $(x_{10},x_{20})$ has a unique continuation in $\Beh_{\Sys_{2}\times\Sys_{c_2}}$.  Furthermore it holds that $\Pi_\Y(\Beh_{\Sys_{2}\times \Sys_{c_2}})\subseteq \Pi_\Y(\Beh_{\Sys_1\times \Sys_{c_1}})$.\qed
\end{pf}
The design of the controller $\Sys_{c_2}$ that achieves exact control refinement for $\Sys_{c_1}$ is similar to that in~\citep{tabuada}, which also holds in the behavioural framework.

\subsection{Driving variable systems}
Since it is difficult to control and analyse a DS directly, we develop a transformation to a system representation that is in non-singular DS form and is driven by an auxiliary input. We refer to this non-singular DS as the  driving variable (DV) system~\citep{weiland1991theory}. We investigate whether the DS and the obtained DV system are bisimilar and behaviourally equivalent. Let us first introduce with a simple example the apparent non-determinism or anti-causality in the DS. Later-on, we show the connections between a DS and its related DV system.
\begin{example}\label{ex:case1}
\it
Consider the DS with dynamics $(E,A,B,C)$ defined as 
\begin{equation}\label{eq:case1}
E=\begin{bsmallmatrix}
1&0&0\\0&0&1\\0&0&0
\end{bsmallmatrix},A=\begin{bsmallmatrix}
-1&0&0\\0&1&0\\0&0&1
\end{bsmallmatrix},B=\begin{bsmallmatrix}
1\\1\\1
\end{bsmallmatrix},C=\begin{bsmallmatrix}
0\\0.2\\0.5
\end{bsmallmatrix}^T,
\end{equation}

and $x(t)=\begin{bmatrix}
x_1(t)&x_2(t)&x_3(t)
\end{bmatrix}^T$. In this case, the input $u(t)=-x_3(t)$ is constrained by the third state component. 
Now the state trajectories of \eqref{eq:case1} can be found as follows: 
\begin{itemize}
\item for a given input sequence $u:\N\rightarrow \A$, we  have $x_2(t)=-u(t)-u(t+1)$, and thus we can use this anti-causal relation of the DS to find the corresponding state trajectories;  
\item alternatively, we can allow the next state $x_2(t+1)$ to be freely chosen, and for arbitrary state $x_2(t)$, the equations \eqref{eq:case1} impose constraints on the input sequence that is, therefore, no longer free as $u(t)=-x_3(t)$.
\end{itemize}
We embrace the latter, non-deterministic interpretation of the DS. 
\end{example}

 This non-determinism can be characterised by  
  introducing an auxiliary driving input of a so-called DV system. 
We reorganise the state evolution of \eqref{eq:DTDS}. For simplicity we omit  the time index in $x(t)$ and $u(t)$ and denote $x(t+1)$ as $x^+$
\begin{align}\label{eq:Mxu}
M\begin{bmatrix}
x^+\\u
\end{bmatrix}=Ax,
\end{align}
where $M=\begin{bmatrix}
E&-B
\end{bmatrix}$. For any $x$, we notice that the pairs $(u,x^+)$ are non-unique due to the non-determinism related to $x^+$. If $M$ has full row rank, then it has a right inverse. This always holds when the DS is reachable (cf. Definition 2-1.1~\citep{dai1989singular}). In that case we can characterise the non-determinism as follows. Let $M^+$ be a right inverse of $M$ such that $MM^+=I$ and 
$N$ be a matrix such that $\im N = \ker
M$ and $N^T N=I$. Then all pairs $(u,x^+)$ that are compatible with state $x$ in \eqref{eq:Mxu} are
parametrised as  
\begin{equation}\label{eq:M+}
\begin{bmatrix}
x^+\\u
\end{bmatrix}=M^+Ax+N s,
\end{equation}
where $s$ is a free variable. We now claim that all transitions $(x,u,x^+)$ in \eqref{eq:M+} for some variable $s$ satisfy \eqref{eq:Mxu}. To see this, multiply $M$ on both sides of \eqref{eq:M+} to  regain \eqref{eq:Mxu}. 
Now  assume that there exists a tuple $(x,u,x^+)$ satisfying \eqref{eq:Mxu} that does not satisfy \eqref{eq:M+}. Then there exists an $s$ and a vector $z\not=0$ that is not an element of the kernel of $M$ and such that the right side of \eqref{eq:M+} becomes $M^+Ax+N s+z$.   
Multiplying again with $M$, we infer that there is an additional non-zero term $Mz$ and that \eqref{eq:Mxu} cannot hold. In conclusion any transition of \eqref{eq:Mxu} is also a transition of \eqref{eq:M+} and vice versa. 


\begin{example}\it[Example \ref{ex:case1}: cont'd]\label{ex:case2}
For the DS of Example \ref{ex:case1}, the related DV system $\DV$ is developed as
\begin{equation}\label{eq:exmpDV}
\begin{aligned}
x(t+1)&=\begin{bsmallmatrix}
-1&0&-1\\0&0&0\\0&1&-1
\end{bsmallmatrix}x(t)+\begin{bsmallmatrix}
0\\-1\\0
\end{bsmallmatrix}s(t)\\
u(t)&=\begin{bsmallmatrix}
0&0&-1\end{bsmallmatrix}x(t)\\
y(t)&=\begin{bsmallmatrix}
0&0.2&0.5
\end{bsmallmatrix}x(t).
\end{aligned}
\end{equation}
As indicated by \eqref{eq:exmpDV}, the input $u(t)$ is a function of the state trajectory. The non-determinism of $x_2(t+1)$ is characterised by $-s(t)$ for which the auxiliary input $s$ can be freely selected.\end{example}
Let us now formalise the notion of a driving variable representation. 
 We associate a driving variable representation with any given DS \eqref{eq:DTDS} by defining a tuple $(A_d,B_d,C_u,D_u,C)$ and setting
\begin{equation}\label{eq:MMA}
\begin{bmatrix}
A_d\\C_u
\end{bmatrix}=M^+A, \begin{bmatrix}
B_d\\D_u
\end{bmatrix}=N,
\end{equation}
where $N\in \R^{(n+p)\times p}$ has orthonormal columns, that is $N^TN=I$. For any given DS, this tuple defines the \emph{driving variable system} $\DV=(\N,\W,\Beh_{\DV})$, which maintains the same set of initial states $\X_0$ and has dynamics  \begin{equation} \label{eq:DVdef}
\begin{aligned}
x(t+1)&=A_dx(t)+B_ds(t)\\
u(t)&=C_ux(t)+D_us(t)\\
y(t)&=Cx(t),
\end{aligned}
\end{equation}  
thereby yielding the initialised behaviour 
\begin{equation}\nonumber
\begin{aligned}
\Beh_{\DV}:=\{w\in\W^{\N}|w=&(u,x,y), \exists s\in  \B^{\N} \\&\mbox{ s.t. } \eqref{eq:DVdef} \mbox{ and  } x_0\in\X_0 \}.
\end{aligned}
\end{equation} 
Next, we propose the following assumption for DS, which will be used in the sequel to develop our main results. \vspace{-3mm}
\subsubsection{Assumption 1.} The given DS $\Sys$ is a dynamical system with dynamics $(E,A,B,C)$ such that $M=\begin{bmatrix}
E&-B
\end{bmatrix}$ has full row rank.

The relationship between a DS and its related DV system is characterised as follows.
\begin{thm}\label{thm:bisimilar}
Let the DS $\Sys$ be given as in \eqref{eq:DTDS} satisfying Assumption 1 and let $\DV=(\N,\W,\Beh_{\DV})$
be defined as in \eqref{eq:DVdef}. Then
\begin{enumerate}
\item $\Sys$ and $\DV$ are bisimilar, that is, $\Sys\cong \DV$,
\item $\Sys$ and $\DV$ have equal behaviour, i.e., $\Beh_{\DV}=\Beh_{\Sys}$,
\item $\Sys$ and $\DV$ have equal output behaviour, that is,
$\Pi_\Y(\Beh_{\Sys})=\Pi_\Y(\Beh_{\DV})
$.
\end{enumerate}


\end{thm} 
\begin{pf}   
For the first statement (1), we define the diagonal relation as $\I:=\{(x,x)\mid x\in \X\}$. Then  $\I$  is a bisimulation relation between $\Sys$ and $\DV$, because by construction their state evolutions can be matched, hence stay in $\I$
; and they share the same output map. In addition, since they have the same set of initial states it follows that 
$\Sys\cong \DV$. \\
The second part (2) follows immediately from the derivation of $\DV$, because by construction all the transitions 
in $\Sys$ can be matched by those of $\DV$ and vice versa, in addition, they have the same output map. Hence, they share the same signal space $(\A\times \X\times \Y)$ and we can conclude that $\Sys$ and $\DV$ have equal behaviour.  \\
Additionally, we have that  (2) implies (3); via Proposition \ref{Def:behimplication} also (1) implies (3). \qed 
\end{pf}

\subsection{Main result: exact control refinement for DS}

Based on the results developed in the previous subsections, we now derive the solution to the exact control refinement problem in Problem 1. More precisely, subject to the assumption that there exists a simulation relation $\re$ from $\Sys_a$ to $\Sys$, for which in addition  holds that $\forall x_{0}\in \X_{0}, \exists x_{a0}\in \X_{a0}$ s.t. $(x_{a0},x_{0})\in \re$, we show that for any $\Sys_{c_a}\in \C(\Sys_a)$, there exists a controller $\Sys_{c}$ for $\Sys$  that refines $\Sys_{c_a}$  
such that $\Sys_{c}\in \C(\Sys)$ and $  \Pi_\Y(\Beh_{\Sys\times \Sys_c})\subseteq \Pi_\Y(\Beh_{\Sys_a\times \Sys_{c_a}})$. 

In the case of Assumption 1, we  construct DV systems $\DV$ and $\DVa$ for the respective DS systems $\Sys$ and $\Sys_a$ as a first step.
For these systems, we develop the following results on exact control refinement:
\begin{enumerate}[i)]
\item The exact control refinement for the DV systems: 
\begin{align*}&\forall\DVa^c\in \C(\DVa), \exists\DV^c\in \C(\DV),\mbox{ s.t.}\\&\hspace{1.6cm}\Pi_\Y\big(\Beh_{\DV\times \DV^c}\big)\subseteq \Pi_\Y\big(\Beh_{\DVa\times \DVa^c}\big);\end{align*}
\item The exact control refinement from $\Sys_a$ to $\DVa$:
 \begin{align*}&\forall\Sys_{c_a}\in \C(\Sys_a),\exists \DVa^c\in \C(\DVa),\mbox{ s.t.}\\&\hspace{2cm}\Pi_\Y\big(\Beh_{\Sys_a\times \Sys_{c_a}}\big)= \Pi_\Y\big(\Beh_{\DVa\times \DVa^c}\big);\end{align*}

\item  The exact control refinement from 
 $\DV$ to $\Sys$:
\begin{align*}&\forall\DV^c\in \C(\DV), \exists \Sys_c\in \C(\Sys),\mbox{ s.t.}\\&\hspace{2.4cm}\Pi_\Y\left(\Beh_{\DV\times \DV^c}\right)= \Pi_\Y\left(\Beh_{\Sys\times \Sys_c}\right).\end{align*}
\end{enumerate}
It will be shown that the combination of the elements i)--iii) also implies the construction of the exact control refinement  for the concrete and abstract DS. 

\subsubsection*{i) Exact control refinement for the DV systems.} 
 
\begin{figure}[!h]
\centering
\includegraphics[width=0.95\columnwidth]{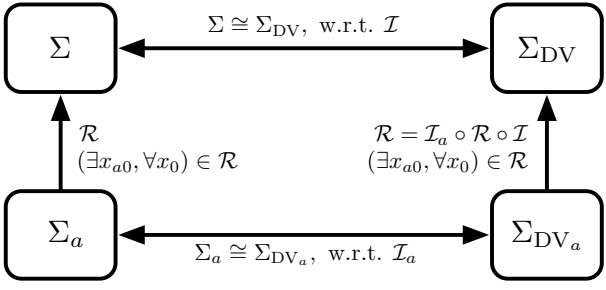}
\caption{Connection between DS and DV systems for the exact control refinement.}
\label{newstructure}
\end{figure}

From Theorem \ref{thm:bisimilar}, we know that $\Sys\cong \DV$ and $\Sys_a\cong\DVa$  with  respective diagonal relations $\I:=\{(x,x){\mid} x\in \X\}$ and $\I_a:=\{(x_a,x_a){\mid} x_a\in \X_a\}$. Hence as depicted in Fig. \ref{newstructure} and based on the transitivity of simulation relations,  we also derive that $\re$ is a simulation relation from $\DVa$ to $\DV$.  

Since the DV systems $\DV$ and $\DVa$ share the same initial states as the respective DS $\Sys$ and $\Sys_a$, it also holds that   
 $\forall x_{0}\in \X_{0}, \exists x_{a0}\in \X_{a0}$ s.t. $(x_{a0},x_{0})\in \re$. According to Theorem \ref{standarddaeexact}, we know that we can do exact control refinement, that is, we have shown \begin{align*}&\forall\DVa^c\in \C(\DVa), \exists\DV^c\in \C(\DV),\mbox{ s.t.}\\&\hspace{1.6cm}\Pi_\Y\big(\Beh_{\DV\times \DV^c}\big)\subseteq \Pi_\Y\big(\Beh_{\DVa\times \DVa^c}\big).\end{align*}   
 

 
\subsubsection{ii) Exact control refinement from $\Sys_a$ to $\DVa$.}
 Denote with $\DVa$ the abstract DV system related to $\Sys_a$, with dynamics $(A_{da},B_{da},C_{u_a},D_{u_a},C_a)$ and initialised with $\X_{a0}$. We first derive the static function $\Sa$ mapping transitions of $\Sys_a$ to the auxiliary input $s_a$ of $\DVa$. From the definition of DV systems, we can also derive the transitions of $\DVa$ indexed with $a$, which is similar to the derivation of \eqref{eq:M+}.
\begin{equation}\label{eq:Ma+}
\begin{bmatrix}
x_a^+\\u_a
\end{bmatrix}=M_a^+A_ax_a+N_a s_a.
\end{equation}
 Multiplying $N_a^T$ on both sides of \eqref{eq:Ma+}, $\Sa$ is derived as 
\begin{equation}\label{eq:S_a}
\Sa:s_a=\Sa(x_a^+,u_a,x_a)=N_a^T \begin{bmatrix}
x_a^+\\u_a
\end{bmatrix}-N_a^T M_a^+A_ax_a.
\end{equation}
$\Sa$ maps the state evolutions of $\Sys_a\times_{w_a} \Sys_{c_a}$ to the auxiliary input $s_a$ for $\DVa$, where $w_a=(u_a,x_a)$. Now, we consider the exact control refinement from the abstract DS to the abstract DV system. 
\begin{thm}\label{thm:abstract_refine}
Let $\Sys_a$ be the abstract DS with dynamics $(E_a, A_a, B_a, C_a)$ satisfying the condition of Assumption 1 and let $\DVa$ be its related DV system with dynamics $(A_{da},B_{da},C_{u_a},D_{u_a},C_a)$ such that both systems are initialised with $\X_{a0}$.
 Then, for any $\Sys_{c_a}\in \C(\Sys_a)$, there exists a controller $\DVa^c\in\C( \DVa)$ that is an exact control refinement for $\Sys_{c_a}$ as defined in Definition \ref{def:exact_refine} with 
\[\Pi_\Y\big(\Beh_{\Sys_a\times \Sys_{c_a}}\big)= \Pi_\Y\big(\Beh_{\DVa\times \DVa^c}\big).\]
\end{thm}
\begin{pf}
Denote with $x_a$ and  $x^d_a$ the state variables of $\Sys_a$ and $\DVa$, respectively.
Next, we construct the controller $\DVa^c$ that achieves exact control refinement for $\Sys_{c_a}$ as \[\DVa^c:=(\Sys_{a}\times_{w_a}\Sys_{c_a})\times_{w_a} \Sys_{\Sa},\] where $w_a=(u_a,x_a)$ and where $\Sys_\Sa:=(\N,\W,\Beh_\Sa)$ is a dynamical system
with 
\begin{equation}\nonumber
\begin{aligned}\Beh_\Sa:=\{(x_a,u_a,x_a^d,s_a)\in \W{\mid} x_{a0}&=x_{a0}^d  \mbox{ and }\\ &s_a=\Sa(x_a^+,u_a,x_a)\}.\end{aligned}
\end{equation}
The dynamical system $\DVa^c$ is a well-posed controller for $\DVa$ with $\DVa\times_{w_a^d}\DVa^c$ sharing $w_a^d=(s_a,x_a^d)$. Denote with $\Beh_{\DVa\times\DVa^c}$ the behaviour of the controlled system. By construction, we know that the set of the behaviour is non-empty and there is a unique continuation for any $x_{a0}^d\in \X_{a0}$. Further based on the construction of $\Sys_{\mathcal{S}_a}$, the behaviour is such that $x_a^d(t)=x_a(t), \forall t\in\mathbb N$.  Additionally, since $\Sys_a$ and $\DVa$ share the same set of initial states $\X_{a0}$, it holds that $\Pi_\Y\left(\Beh_{\Sys_a\times \Sys_{c_a}}\right)= \Pi_\Y\left(\Beh_{\DVa\times \DVa^c}\right).$ 
\qed\end{pf}

The proof is actually constructive in the design of the controller $\DVa^c$ that achieves exact control refinement for $\Sys_{c_a}$.
\subsubsection{iii) Exact control refinement from 
 $\DV$ to $\Sys$.}
 Now, we consider the exact control refinement from $\DV$ to $\Sys$.
 Suppose we are given a well-posed controller $\DV^c$ for $\DV$, which shares the free variable $s$ and the state variable $x$ with $\DV$. 
 We want to design a well-posed controller for $\Sys$ over $w=(u,x)$, 
for which we consider the dynamical system $\SysX=(\mathbb N, \mathbb W, \Beh)$ over the signal space $\mathbb W=\A\times \X\times \mathbb S$, the behaviour of which can be defined by 
\begin{equation}\label{eq:refinelaw}
\begin{aligned}
B_{d}^Tx(t+1)&=B_{d}^TA_{d}x(t)+B_{d}^TB_{d}s(t)\\
u(t)&=C_ux(t)+D_us(t).
\end{aligned}
\end{equation} \vspace{-2mm}

Then the dynamics of the interconnected system $\Sys\times_{w} \SysX$ as a function of $x$ and $s$ is derived as 
\begin{equation}\label{eq:SX}
\begin{bmatrix}E\\B_d^T\end{bmatrix}x(t+1)=\begin{bmatrix}A+BC_u\\B_d^TA_d\end{bmatrix}x(t)+\begin{bmatrix}BD_u\\B_d^TB_d\end{bmatrix}s(t).
\end{equation}

Note that $A+BC_u=EA_d$ and $BD_u=EB_d$ by multiplying $M=\begin{bmatrix}
E&-B
\end{bmatrix}
$ on the left-hand side of the two equations in \eqref{eq:MMA}. Therefore, \eqref{eq:SX} is simplified to
\begin{equation}\label{eq:interconnected}
\begin{bmatrix}E\\B_d^T\end{bmatrix}x(t+1)=\begin{bmatrix}E\\B_d^T\end{bmatrix}A_dx(t)+\begin{bmatrix}E\\B_d^T\end{bmatrix}B_ds(t).
\end{equation}
Furthermore $\begin{bmatrix}E^T&B_d\end{bmatrix}^T$ has full column rank because the matrix $\begin{bmatrix}M^T&N\end{bmatrix}^T$ is square and has full rank. Hence $\begin{bmatrix}E^T&B_d\end{bmatrix}^T$ has a left inverse and the dynamics of   $\Sys\times_w \SysX$ in \eqref{eq:interconnected} can be simplified as
\begin{equation}\nonumber
x(t+1)=A_dx(t)+B_ds(t),
\end{equation}
which is exactly the same as the state evolutions of $\DV$ as shown in \eqref{eq:DVdef}. Next we construct $\Sys_c:=\SysX \times_{w^d} \Sys_{DV}^c$ with $w^d=(s,x^d)$ and it is a well-posed controller for $\Sys$. This allows us to state the following theorem regarding the control refinement from $\DV$ to $\Sys$.

\begin{thm}\label{thm:concrete_refine} 
Let $\Sys$ be the concrete DS with dynamics $(E, A, B, C)$ satisfying Assumption 1 and let $\DV$ be its related DV system with dynamics $(A_{d},B_{d},C_{u},D_{u},C)$ such that both systems are initialised with $\X_{0}$. Then, for any $\DV^c\in \C(\DV)$, there exists a controller $\Sys_c\in\C( \Sys)$ that is an exact control refinement for $\DV^c$ as defined in Definition \ref{def:exact_refine} with \[\Pi_\Y\left(\Beh_{\DV\times \DV^c}\right)= \Pi_\Y\left(\Beh_{\Sys\times \Sys_c}\right).\]
\end{thm}

\begin{pf}\label{pf:concrete_refine}
Denote with $x$ and  $x^d$ the state variables of the $\Sys$ and $\DV$, respectively. Next, we construct the controller $\Sys_c$ that achieves exact control refinement for $\DV^c$ as\[\Sys_c:=\SysX \times_{w^d} \DV^c,\]
where $w^d=(s,x^d)$ and the dynamics of $\SysX$ is defined as \eqref{eq:refinelaw}. Then, we can show that the dynamical system 
$\Sys_{c}$ is a well-posed controller for $\Sys$.
Based on the analysis of \eqref{eq:interconnected}, it is shown that $\Sys\times_w \SysX=\DV$ with $w=(u,x)$, then we can derive $\Sys\times_w \Sys_c=\DV\times_{w^d} \DV^c$. Therefore, we can conclude $\Sys_c\in \C(\Sys)$ with $\Pi_\Y\big(\Beh_{\DV\times \DV^c}\big)= \Pi_\Y\big(\Beh_{\Sys\times \Sys_c}\big)$ immediately follows from $\DV^c\in \C(\DV)$.  \qed
\end{pf}
\subsubsection{Exact control refinement for descriptor systems.}
We can now argue that there exists exact control refinement 
 from $\Sys_a$ to $\Sys$, as stated in the following result.
\begin{thm}\label{thm:solution1}
Consider two DS $\Sys_a$ (abstract, initialised with $\X_{a0}$) and $\Sys$ (concrete, initialised with $\X_{0}$) satisfying Assumption 1 and let $\re$ be a simulation relation from $\Sys_a$ to $\Sys$, for which in addition holds that $\forall x_{0}\in \X_{0}, \exists x_{a0}\in \X_{a0}$ s.t. $(x_{a0},x_{0})\in \re$. Then, for any $\Sys_{c_a}\in \C(\Sys_a)$, there exists a controller $\Sys_c\in\C(\Sys)$ such that \[
\Pi_\Y\left(\Beh_{\Sys\times \Sys_{c}}\right)\subseteq \Pi_\Y\left(\Beh_{\Sys_a\times \Sys_{c_a}}\right).
\]
\end{thm}

\begin{pf}\label{pf:solution1}
Based on Assumption 1, we first construct $\DV$ and $\DVa$. Then to prove this we need to  construct the exact  control refinement.
%
This can be done based on the subsequent control refinements given in Theorem \ref{thm:abstract_refine}, Theorem \ref{standarddaeexact} and  Theorem  \ref{thm:concrete_refine}. \qed \end{pf}

Theorem \ref{thm:solution1} claims the existence of such controller $\Sys_c$ that achieves exact control refinement for $\Sys_{c_a}$. More precisely, we have shown in the proof that the refined controller $\Sys_c$ is constructive, which provides the solution to Problem 1.

To elucidate how such an exact control refinement is constructed, we consider the following example.

\begin{example}\it[Example \ref{ex:case1},\ref{ex:case2}: cont'd]\label{ex:case3}
Consider the DS of Example \ref{ex:case1} and its related DV system (cf. Example \ref{ex:case2}) such that both systems are initialised with $\X_0=\{x_0\mid x_0\in [-1,1]^3 \subset \R^3\}$.
According to Silverman-Ho algorithm~\citep{dai1989singular}, we can select an abstract DS $\Sys_a=(E_a,A_a,B_a,C_a)$ that is the minimal realisation of $\Sys$ and is initialised with $\X_{a0}=\R^2$, in addition 
\begin{equation}\nonumber
E_a=\begin{bsmallmatrix}
0&0\\1&0
\end{bsmallmatrix},A_a=\begin{bsmallmatrix}
1&0\\0&1
\end{bsmallmatrix},B_a=\begin{bsmallmatrix}
1\\0
\end{bsmallmatrix},C_a=\begin{bsmallmatrix}
0.7\\0.2
\end{bsmallmatrix}^T.
\end{equation}
Similarly, the related DV system $\DVa$ of $\Sys_a$ is given as
\begin{equation}\label{eq:ex4DVa}
\begin{aligned}
x_a(t+1)&=\begin{bsmallmatrix}
0&1\\0&0
\end{bsmallmatrix}x_a(t)+\begin{bsmallmatrix}
0\\-1
\end{bsmallmatrix}s_a(t)\\
u_a(t)&=\begin{bsmallmatrix}
-1&0\end{bsmallmatrix}x_a(t)\\
y_a(t)&=\begin{bsmallmatrix}
0.7&0.2
\end{bsmallmatrix}x_a(t).
\end{aligned}
\end{equation}
Subsequently, 
\begin{equation}\nonumber
\mathcal{R}:=\{(x_a,x)\mid x_a=\mathcal{H}x, x_a\in \X_a, x\in \X\}
\end{equation}
is a simulation relation from $\Sys_a$ to $\Sys$ with
\begin{equation}\nonumber
\mathcal{H}=\begin{bsmallmatrix}
0&0&1\\0&1&-1
\end{bsmallmatrix}.
\end{equation}
This can be proved through verifying the two properties of Definition \ref{Def:simrelation}. In addition, the condition $\forall x_{0}\in \X_{0}, \exists x_{a0}\in \X_{a0}$ s.t. $(x_{a0},x_{0})\in \re$ holds. According to Theorem \ref{thm:solution1}, we can refine any $\Sys_{c_a}\in \C(\Sys_a)$ to attain a well-posed controller $\Sys_c$ for $\Sys$ that solves Problem 1 as follows: Define $\Sys_{c_a}\in \mathfrak{C}(\Sys_a)$ with dynamics as
\begin{equation}\nonumber
\begin{bsmallmatrix}
1&1
\end{bsmallmatrix}x_a(t+1)=\begin{bsmallmatrix}
0.5&0.5
\end{bsmallmatrix}x_a(t)+u_a(t).
\end{equation}
The controlled system $\Sys_a\times_{w_a} \Sys_{c_a}$ is derived as
\begin{equation}\nonumber
\begin{aligned}
x_a(t+1)&=\begin{bsmallmatrix}
0&1\\-0.5&-0.5
\end{bsmallmatrix}x_a(t)\\
y_a(t)&=\begin{bsmallmatrix}
0.7&0.2
\end{bsmallmatrix}x_a(t),
\end{aligned}
\end{equation}
 with $w_a=(u_a,x_a)$ and $u_a(t)=\begin{bsmallmatrix}
-1&0\end{bsmallmatrix}x_a(t)$. Then $\Sys_a\times_{w_a} \Sys_{c_a}$ is stable. According to Theorem \ref{thm:abstract_refine}, we derive the map $\mathcal{S}_a$ for $\DVa$ as $s_a(t)=\begin{bsmallmatrix}0&-1\end{bsmallmatrix}x_a(t+1)=\begin{bsmallmatrix}0.5&0.5\end{bsmallmatrix}x_a(t).$ Next, the related interface from $\DVa$ to $\DV$ is developed as $s(t)=s_a(t)-\begin{bsmallmatrix}0&1&-1\end{bsmallmatrix}x(t).$ According to Theorem \ref{thm:concrete_refine}, we derive the well-posed controller $\Sys_c$ as
\begin{equation}\nonumber
\begin{aligned}
\begin{bsmallmatrix}
0&-1&0
\end{bsmallmatrix}x(t+1)&=\begin{bsmallmatrix}
0&-1&1
\end{bsmallmatrix}x(t)+\begin{bsmallmatrix}
0.5&0.5
\end{bsmallmatrix}x_a(t)\\
u(t)&=\begin{bsmallmatrix}
0&0&-1
\end{bsmallmatrix}x(t),
\end{aligned}
\end{equation}
and the interconnected system $\Sys\times_w \Sys_c$ with $w=(u,x)$, is derived as
\begin{equation}\nonumber
\begin{aligned}
x(t+1)&=\begin{bsmallmatrix}
1&0&1\\0&1&-1\\0&1&-1
\end{bsmallmatrix}x(t)+\begin{bsmallmatrix}
0&0\\-0.5&-0.5\\0&0
\end{bsmallmatrix}x_a(t)\\
y(t)&=\begin{bsmallmatrix}
0&0.2&0.5
\end{bsmallmatrix}x(t).
\end{aligned}
\end{equation}
Since $(x_a,x)\in \re$, that is $x_a=\mathcal{H}x$, $\Sys\times_w \Sys_c$ can be simplified by replacing $x_a(t)$:
\begin{equation}\nonumber
\begin{aligned}
x(t+1)&=\begin{bsmallmatrix}
1&0&1\\0&0.5&-1\\0&1&-1
\end{bsmallmatrix}x(t)\\
y(t)&=\begin{bsmallmatrix}
0&0.2&0.5
\end{bsmallmatrix}x(t).
\end{aligned}
\end{equation}
Finally, $\Sys_c\in \C(\Sys)$ and \(
\Pi_\Y\left(\Beh_{\Sys\times \Sys_{c}}\right)\subseteq \Pi_\Y\left(\Beh_{\Sys_a\times \Sys_{c_a}}\right)
\) are achieved.
%

\end{example}


 \section{Conclusion}%
\label{sec:Conclusion}
In this paper, we 
 have developed a control refinement procedure for discrete-time descriptor systems that is largely based on the behavioural theory of dynamical systems and the theory of simulation relations among dynamical systems. Our main results provide complete solutions of the control refinement problem for this class of discrete-time systems.  
 
The exact control refinement that has been developed in this work also opens the possibilities for \emph{approximate} control refinement notions, to be coupled with approximate similarity relations: these promise to leverage general model reduction techniques and to provide more freedom for the analysis and control of descriptor systems.  

The future research includes a comparison of the control refinement approach for descriptor systems to results in perturbation theory, 
as well as control refinement for nonlinear descriptor systems.

%
%

\bibliography{ifacconf}             

\end{document}